\documentstyle[sprocl,epsfig]{article}

\def\ra{\rightarrow}

\def\be{\begin{equation}}
\def\ee{\end{equation}}
\def\bea{\begin{eqnarray}}
\def\eea{\end{eqnarray}}

\def\m0{M_0}
\def\mhf{M_{1/2}}
\def\neuto{\tilde{\chi}_1^0}
\def\mneuto{m_{\tilde{\chi}_1^0}}
\def\neutt{\tilde{\chi}_2^0}
\def\mneutth{m_{\tilde{\chi}_3^0}}

\def\chargop{\tilde{\chi}_1^+}

\def\stau{\tilde{\tau}_1}

\def\mt{m_{t}}
\def\ma{M_{A}}

\newcommand{\beqn}{\begin{eqnarray}}
\newcommand{\eeqn}{\end{eqnarray}}

\begin{document}

\title{UNCERTAINTIES IN THE PREDICTION OF THE RELIC DENSITY OF SUPERSYMMETRIC  DARK MATTER\\}

\author{BENJAMIN ALLANACH, GENEVIEVE BELANGER, FAWZI BOUDJEMA}

\address{Laboratoire de Physique Th\'eorique {\large
LAPTH},\\
 Chemin de Bellevue, B.P. 110, F-74941 Annecy-le-Vieux,
Cedex, France.}

\author{ALEXANDER PUKHOV}
\address{{\it  Skobeltsyn Institute of Nuclear Physics,
Moscow State University} \\ {\it Moscow 119992,
Russia }\\}


\maketitle\abstracts{
We investigate how well the relic density of dark matter can be predicted in mSUGRA. We determine the parameters to which the relic density is most sensitive and quantify the collider accuracy needed to match the accuracy of WMAP and PLANCK.
}

One of the very attractive features of the minimal supersymmetric
standard model is that it provides a natural candidate for cold
dark matter. Cosmology, and in particular the measurement of the
relic density of dark matter then put some strong constraints on
the supersymmetric model. For example, WMAP [1] which at $2\sigma$
constrains the relic density in the range $.094<\Omega h^2<.129$ 
points to very specific scenarios [2]. Within the
context of mSUGRA models, the favoured  scenarios are
 slepton coannihilation, heavy Higgs annihilation
 and Higgsino LSP. The latter occurs in the focus point region with
heavy scalars.
 Predictions for the relic density
 depend sensitively on the details of the supersymmetric models but
also on some assumptions on the cosmological model. 
Assuming the LHC will discover
supersymmetry, the question  then is whether one can make
precise predictions on the value of the relic density of
dark matter, hence confront the cosmological model.  

In the following we investigate to which extent precision measurements at a
linear collider can match the accuracy on the relic density of
dark matter as measured my WMAP (10\% level) or as expected from  PLANCK (2\% level).
To estimate this we will concentrate on a few observables, the
ones that are relevant in the rather fine-tuned mSUGRA scenarios allowed by 
WMAP without assuming mSUGRA when analysing data [3].
 The relic density is computed with {\tt micrOMEGAs1.3} [4] and the evaluation of the supersymmetric spectrum relies on {\tt SoftSUSY}1.8.7 [5].

\section{Coannihilation}

\begin{figure}[ht] {
 \unitlength=1.1in
\begin{picture}(0.2,1.6)
\put(0.1,0){\epsfig{file=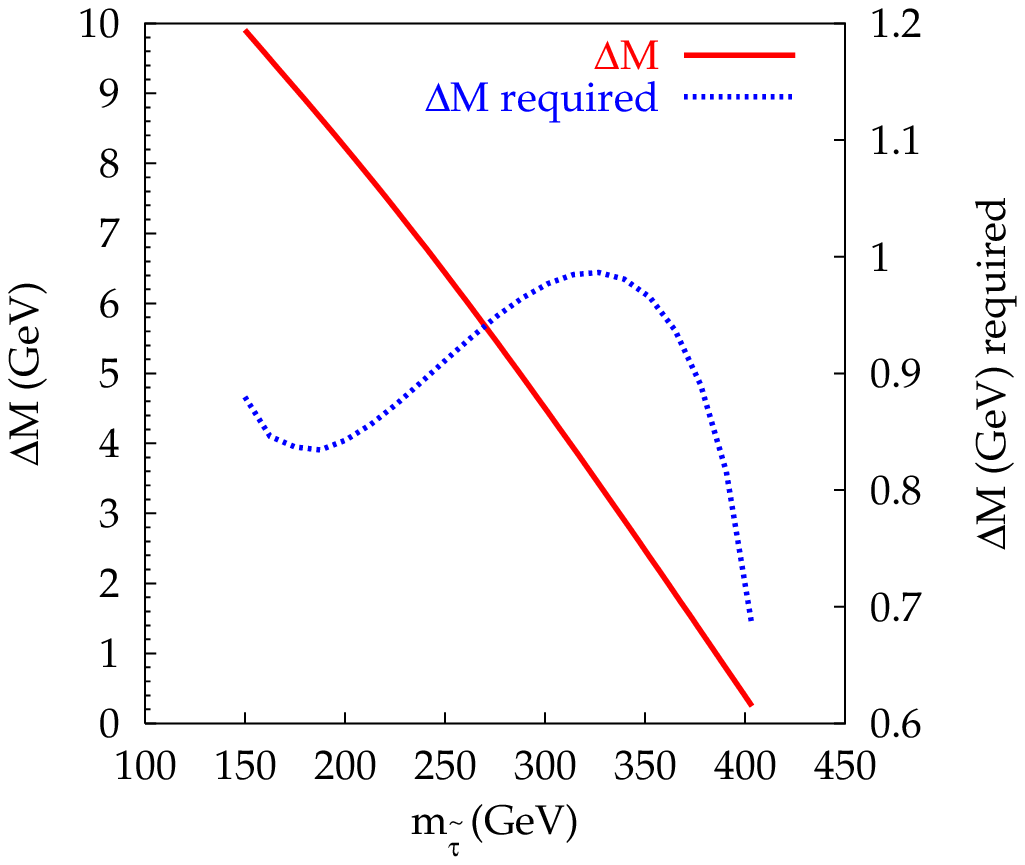,width=7.cm}}
\end{picture}
\begin{picture}(0.2,1.6)
\put(2.1,0){\epsfig{file=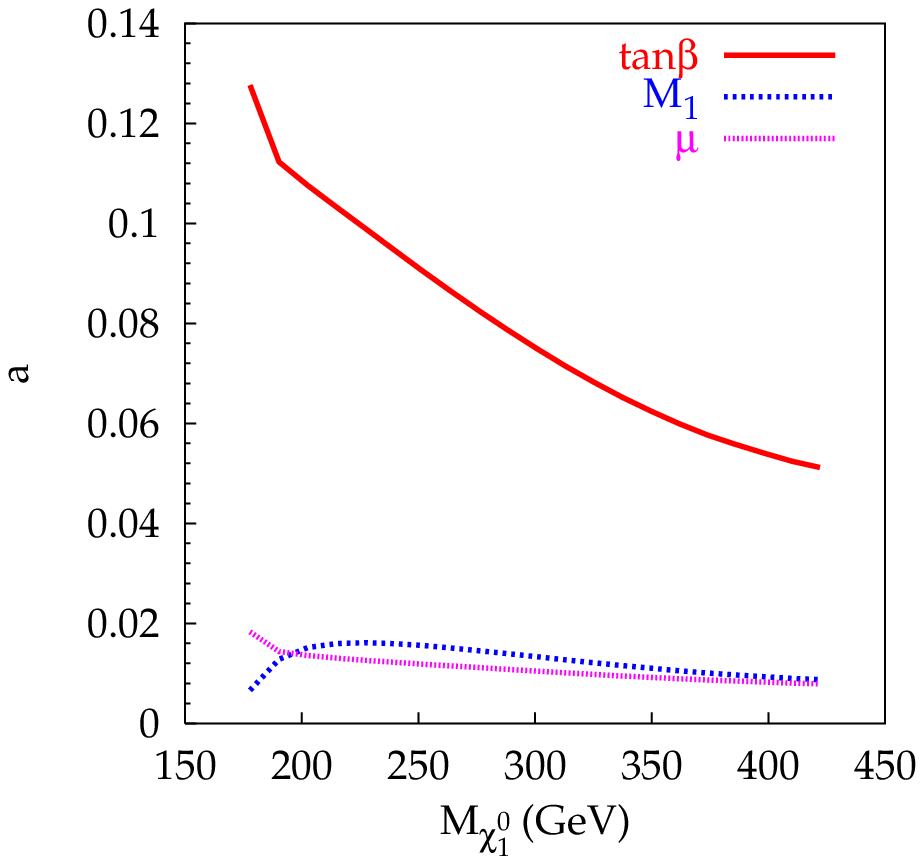,width=7.cm}}
\end{picture}
\caption{\it \small(a) Mass difference $\Delta M_{\stau\neuto}$ along slope S1 and precision required  to match WMAP accuracy
     (b) Sensitivity to  $\mu,M_1$ and $\tan\beta$  along slope S2. }
  \label{fig:coan}}
\end{figure}

In mSUGRA at small $\m0$ there exists a region with almost degenerate $\stau-\neuto$. The contribution of coannihilation channels to the effective annihilation cross section, in particular $\neuto\stau\ra \tau \gamma$ or $\stau\stau\ra \tau\tau$, is essential in bringing the relic density in the desired range. 
 In computing the effective
annihilation cross section, coannihilation processes are
suppressed by a Boltzmann factor $\propto exp^{-\Delta M/T_f}$
where $\Delta M$ is the mass difference between the NLSP and the
LSP and $T_f$ the decoupling temperature. One expects $\Omega h^2$ to be very sensitive to this mass difference.
Following Ref.~[6], we take a slope ''S1" in parameter space 
with $\mu>0,\tan\beta=10$, $A_0=0$ and
\beqn
\m0=20.3333+0.134\mhf+5.0667\times 10^{-5}(\mhf)^2
\eeqn
where masses are  in GeV. 
Along this slope ($\mhf=350-950$~GeV), the relic density is in rough agreement with the WMAP range.
Fig.~\ref{fig:coan} shows that the mass difference between $\stau$ and $\neuto$ varies from 12 GeV to 300 MeV as one increases the $\stau$ mass. 
The mass difference between the LSP and the sleptons of the first
two generations  can be much larger.
To estimate the sensitivity of $\Omega h^2$  on the mass difference we have, for each point on the slope,  varied the mass of the $\stau$ while keeping all other parameters fixed.   
 We find that for the staus accessible at  LC500, the mass difference must be measured to 
 0.9(0.2) GeV to match the WMAP (PLANCK) accuracy as displayed in  Fig.~\ref{fig:coan}a.
Recently two groups have studied the
detectability of  nearly degenerate $\stau-\neuto$ as well as the
feasability of a precise measurement of the $\stau$ mass at linear colliders [7].
The results indicate  that the
level of accuracy needed to match PLANCK might just be reached.

\section{Large $\tan\beta$}
In the Higgs funnel region, rapid  annihilation of neutralinos proceed via the s-channel exchange of a heavy pseudoscalar Higgs. This region is found at large values of $\tan\beta$ and is accessible to LC500 only for low values of $\mhf$. 
Within mSUGRA the relic density is sensitive to both $m_b(m_b)$  as well as the top quark mass since these strongly affect the value of the pseudoscalar mass.
 One can reduce significantly this
theoretical uncertainty  by measuring directly some
of the physical parameters. 
 The crucial parameters in this scenario  are the masses $\mneuto,\ma$.
As an example we consider a mSUGRA model 
 with $\tan\beta=50$, $\mhf=380$GeV, $\m0=399$GeV,  $A_0=0,\mu>0$, and then vary either $\mneuto$ (through $M_1$) or $\ma$ while keeping {\it all}  other parameters fixed. We find that to match WMAP (PLANCK) accuracy on $\Omega h^2$,
$\ma$ and $\mneuto$ need to be measured with a precision of $1\%(0.2\%)$.
Note that the LHC can measure precisely $\ma$ provided
the $A\ra \mu^+\mu^-$ channel can be used [8].

\section{Higgsino LSP}
In mSUGRA, a LSP with a significant Higgsino component can be found at high values of $\m0$ near the boundary of viable electroweak symmetry breaking 
where the value of $\mu$ drops rapidly. This is the focus point region.
 The  enhanced coupling of the Higgsino pairs to the Z and to the Higgses then makes for efficient annihilation into gauge bosons or $t\bar{t}$ pairs. 
Coannihilation channels
with  $\neutt/\chargop$ also contribute to the relic density.

The position of the focus point region is  extremely sensitive to the
value of the  running top quark mass 
that enter the spectrum calculator codes [3,9].  For $\mt=175$~GeV, the region where $\mu\approx M_1$ and where the LSP has some  Higgsino  component is found generally for high values of $\tan\beta$.  We will therefore consider a model with  $\tan\beta=50,\mu>0,A_0=0$, and define the slope  S2 where the value of $\Omega h^2$ is in rough agreement with WMAP:
 \begin{equation}
m_0 = 3019.85 +  2.6928M_{1/2} -1.01648
\times 10^{-4} \left(M_{1/2}\right)^2.
\end{equation}
We have estimated the sensitivity   of the relic density prediction to  $m_t$ along S2 in mSUGRA. Reaching  WMAP precision requires knowing $m_t$ to 20 MeV. 
This is too demanding considering that
at a linear colider one expects a precision of 100MeV on $m_t$ when including  theoretical uncertainties.

The situation improves significantly if one does not rely on mSUGRA and uses the information that could be provided by colliders, the $\mt$ dependence drops to the few percent level. 
Here  the Higgsino fraction determined by $\mu, M_1$ is expected to be the crucial parameter. Along the slope S2, we vary either $\mu$ or $M_1$  and recalculate the whole spectrum. We then estimate the accuracy $a$ required on each parameter to  get a $10\%$ shift in $\Omega h^2$. As  displayed in
Fig.~\ref{fig:coan}b, a precision  of $1-1.5\%$ is required, while matching PLANCK accuracy requires roughly  an accuracy of $0.2-0.3\%$.  Since $\mneuto$ and $\mneutth$ are directly related to $\mu$ and $M_1$ a precise measurement of these two masses would allow to reconstruct the necessary parameters. The situation for $\tan\beta$ is more problematic as even the precision required for WMAP (10\%) might be difficult to reach.
Nevertheless a realistic study of the achievable accuracy  in this scenario  at a linear collider still needs to be performed.

\section {Conclusion}

 The precise knowledge of the cross-sections necessary to make an accurate prediction of the relic density of dark matter rests on a precise knowledge of the physical parameters of the MSSM. Although the relic density calculation often involves a large number of processes, 
 within the scenarios that are favoured by WMAP, only  a few parameters need to be measured with very high precision. 
Furthermore, using the collider data rather than relying on some theoretical prejudice, considerably improves the precision of the prediction of $\Omega h^2$.

\end{document}